\documentstyle[aps,prb,floats,epsf,epsfig,amstex]{revtex}

\begin{document}
\newcommand{\eqbeg}{\begin{equation}}
\newcommand{\eqend}{\end{equation}}
\newcommand{\shat}{\hat{S}}
\newcommand{\ddt}{\frac{d}{dt}}
\newcommand{\thalf}{\mbox{T}_{1/2}}

\wideabs{

\title{Correlated Fluctuations between Luminescence and Ionization
in Liquid Xenon}

\author{E.~Conti$^{1,2\dagger}$, R.~DeVoe$^1$, G.~Gratta$^1$, T.~Koffas$^1$,
        S.~Waldman$^1$, J.~Wodin$^1$, D.~Akimov$^3$, G.~Bower$^2$,
        M.~Breidenbach$^2$, R.~Conley$^2$, M.~Danilov$^3$,
        Z.~Djurcic$^4$, A.~Dolgolenko$^3$, C.~Hall$^2$,
        A.~Odian$^2$, A.~Piepke$^4$, C.Y.~Prescott$^2$, P.C.~Rowson$^2$,
        K.~Skarpaas$^2$, J-L.~Vuilleumier$^5$,
        K.~Wamba$^2$, O.~Zeldovich$^3$} 

\address{$^1$ Physics
        Department, Stanford University, Stanford CA, USA \\ 
	$^2$ Stanford Linear Accelerator Center, Stanford University,
        Stanford CA, USA \\
	$^3$ Institute for Theoretical and Experimental Physics,
        Moscow, Russia \\ 
	$^4$ Department of Physics and Astronomy, University of
        Alabama, Tuscaloosa AL, USA \\ 
	$^5$ Institut de Physique, Universite de Neuchatel, Neuchatel,
        Switzerland  \\
        $\dagger$ On leave from INFN Padova, Padova, Italy }

\date{\today}

\maketitle

\begin{center}
                       (EXO Collaboration)
\end{center}

\begin{abstract}

  The ionization of liquefied noble gases by radiation is known to be
accompanied by fluctuations much larger than predicted by Poisson
statistics.  We have studied the fluctuations of both scintillation
and ionization in liquid xenon and have measured, for the first time, 
a strong anti-correlation between the two at a microscopic level, with 
coefficient $ -0.80 < \rho_{ep} < -0.60 $.   This provides direct
experimental evidence that electron-ion recombination is partially
responsible for the anomalously large fluctuations and at the same
time allows substantial improvement of calorimetric energy resolution.

\end{abstract}


}


The measurement of ionizing radiation in a liquiefied noble
gas\cite{doke4} such as Ar, Kr, or Xe can be characterized by two
parameters: $W_e$, the mean energy required to create a free electron-ion
pair and the Fano factor, $F_e$, which parameterizes the fluctuations
in the number of ion pairs.\cite{fano1} The Fano factor is defined by
\begin{equation}
        \sigma_e = \sqrt{F_e N_e}\label{eqn:Fano}
\end{equation}
where $\sigma_e^2$ is the variance of the charge expressed in units of
the electron charge, $e$, and $N_e$ is the number of ion pairs.  $F_e=1$
corresponds to Poisson statistics. Fano originally predicted that
the charge fluctuations would be sub-Poissonian, $F_e < 1$, because
the individual ion-pair creation processes are not independent once
the additional constraint $E = N_e W_e$ is included, where $E$ is the
energy deposited by the incident particle. The Fano factor for gaseous noble
elements is relatively well understood, where for example the theory
for argon predicts $F_e = 0.16$ while experiment yields $F_e =
0.20$.\cite{dokepra}
 
The Fano factor for the liquid phase has been far more difficult to
understand, even though the values of $W_e$ are similar to those of the
gas. The pioneering theoretical work of Doke \cite{doke1} in 1976
predicted $F_e \approx 0.05$ for liquid Xe (LXe).  Experiments,
however, showed $F_e\ > 20$, that is, a variance 20 times worse than
Poissonian and 400 times worse than predicted by Fano's original
argument.  This discrepancy has not only raised important issues for
the understanding of the phenomena of energy loss and
conversion, but it is also of interest for experimental physics since
this large Fano factor limits the resolution of calorimeters used in
nuclear and particle physics. 

Luminescence light provides a second process, complementary to
ionization, with which to study energy loss and conversion
phenomena. The properties of luminescence light emitted by ionizing
radiation (scintillation) in liquid Ar, Kr, and Xe detectors have been
extensively studied and have been exploited in
calorimetry\cite{doke4,DokeMasuda}. As described elsewhere\cite{schwentner}, 
scintillation photons are produced by the relaxation of a xenon 
excimer, $\mathtt{Xe_2^*} \rightarrow 2 \mathtt{Xe} + \gamma$.   
The excimer is
formed in two ways; 1) direct excitation by an energetic particle, 2)
recombination of an electron with a xenon molecular ion, $
\mathtt{Xe_2^+ + e} \rightarrow \mathtt{Xe_2^{*}} $ (recombination 
luminescence)\cite{macro_corr}.    The electron-ion
recombination rate depends upon the ionization density and the applied
electric field. The quantities $W_p$ and $F_p$ can be defined in
analogy with the ionization case, where $W_p$ is the mean energy
absorbed per emitted photon and $F_p$ parameterizes their
variance. Fano's theory was not intended to treat photon fluctuations
and in this case $F_p$ is treated as a phenomenological parameter.

\begin{figure}[htb!!]
\begin{center}
\mbox{\epsfig{file=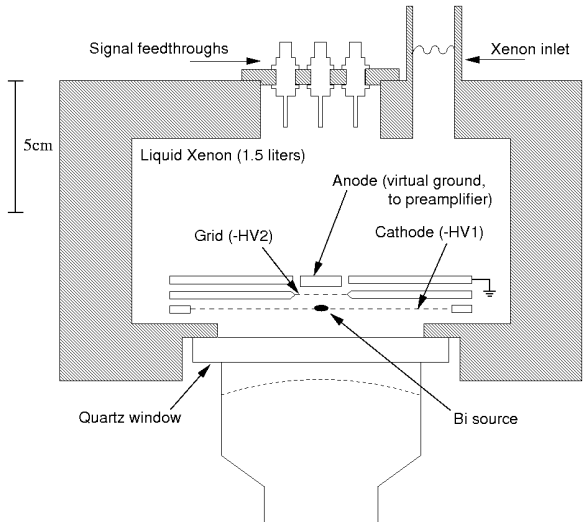,height=6.9cm}}
\vskip 0.1cm
\caption{Schematic drawing of the liquid xenon detector.  As described
in the text, the chamber is cooled by immersion in a refrigerant bath.
The bismuth source is in reality much smaller than indicated in the
figure.}
\label{fig:chamberSchematic}
\end{center}
\end{figure}

We have simultaneously studied both ionization and scintillation in
LXe using a special geometry ionization chamber equipped with a VUV
photomultiplier tube (PMT) capable of detecting the $\simeq 175$~nm
scintillation light from the xenon.  The chamber, shown in
Figure~\ref{fig:chamberSchematic}, has a pancake geometry with a total
volume of $\simeq 1.5$~liters.  It is built out of ultra-high-vacuum
compatible parts, so that it can be baked and treated with hot xenon
before filling to obtain good LXe purity.  The chamber is cooled by
immersion in a bath of HFE7100~\cite{HFE7000} that is in turn cooled
by liquid nitrogen.  The xenon is purified by an
Oxysorb~\cite{Oxysorb} cartridge during transfer from the storage
cylinder to the chamber.  Ionization is collected in a fiducial volume
with radius $\simeq$~10~mm and height $\simeq$~6~mm defined by a small grid
and a large cathode plane.  The grid and cathode are made of an
electroformed nickel mesh with 90\% optical transparency.  Drifted
electrons are collected on a low-capacitance anode surrounded by a
field defining ring held at the same ground potential.  An electron
transparency of 100\% is obtained by maintaining the anode-to-grid field,
($E_{AG}$), double in value with respect to the drift field,
($E_{KG}$)\cite{transparency}, for all the measurements reported here.   
All data was collected at a LXe temperature of $169.2\pm 0.5$~K and 
pressure of $990^{+30}_{-50}$~Torr.   The temperature gradients in 
the chamber were measured to be less than $\pm 0.2$~K.

The electron signal is detected by a low noise charge sensitive
preamplifier\cite{amptek} with a cold input FET placed on the
feedthrough of the chamber.  After subsequent amplification and
filtering, the charge signal $f(t)$ is recorded on a transient
digitizer and fit to a model assuming drift to the anode of a gaussian
ionization distribution represented by the function

\begin{equation}
f(t)=A\left[1+erf\left(\frac{\Delta t}{C}\right)\right]
e^{-\frac{t}{\tau}} + \alpha\label{eqn:fitFunction}
\end{equation}

\noindent where $\Delta t = t-t_0$ and the term $\alpha = D
e^{-\frac{t}{\tau}} + 100 + 0.019t$ accounts for pile-up, ADC offset
and a small integral non-linearity.  We simultaneously fit for the
amplitude $A$, the effective drift-time $t_0$, the rise-time parameter
$C$, the pile-up offset $D$ and the integrator fall-time constant $\tau$. 
The fit fall-time
is $\simeq 50~\mu$s for all the data, as expected.  The parameter $C$
includes information about the multiplicity of the
ionization event and its values range from 0.78~$\mu$s for a field of
4~kV/cm to 1.04~$\mu$s for 0.2~kV/cm.  The decrease of this parameter
with the increase of the electric field is consistent with the
expected change in drift velocity.

The readout is calibrated in units of electron charge by injecting a
signal into the input FET through a calibrated capacitor.  The noise
of the preamplifier was measured with a test signal and found to
follow a gaussian distribution with width $\sigma_N = 381 e$.

The scintillation light is detected through the cathode grid by the
3-inch VUV PMT~\cite{PMT} with a solid angle that varies from 21.5\%
to 26\% of $4\pi$ for signals produced in the region between cathode and grid
active for ionization collection.  The fraction of photoelectrons
extracted from the PMT photocathode per scintillation photon produced
in the LXe varies between 3.0\% and 3.6\% considering mesh and quartz
window transparencies and photocathode quantum efficiency (22\%
including transparency of the PMT window).  The PMT signal is read out
using a 12 bit charge ADC calibrated by measuring the single-photon
peak, which is well resolved from noise.

Data acquisition (DAQ) is triggered when there is a delayed
coincidence between PMT and charge signals.  The delay is required to
account for the electron drift-time and varies with electric
field. Once a coincidence is established, DAQ is inhibited until the
event has been processed and the instruments cleared.  About 25\% of
the collected data is removed off-line by analysis cuts that reject
events that don't fit model~(\ref{eqn:fitFunction}).  These ``bad
events'' are mostly due to multiple ionization deposits from Compton
scattering, deposits at the edge of the ionization fiducial region,
and cosmic ray background.

\begin{figure}[tb!!]
\begin{center}
\mbox{\epsfig{file=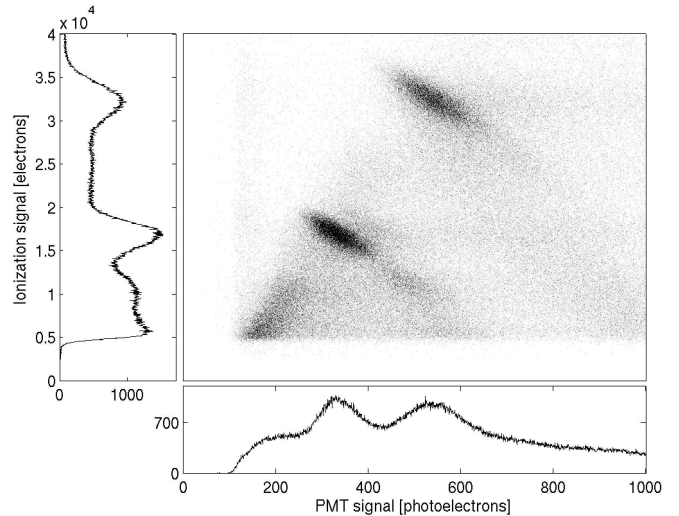,height=6.9cm}}
\vskip 0.1cm
\caption{The two-dimensional scintillation and ionization spectra
recorded at drift field $E_{KG}=4$kV/cm.  The two ``islands'' with
negative correlation coefficient correspond to the two $\gamma$ lines
from the $^{207}$Bi source and their satellite internal conversion
peaks.  The axes are calibrated in terms of absolute numbers of
elementary excitations (ionization electrons and photo-electrons in 
the PMT).}
\label{fig:4kv_correl}
\end{center}
\end{figure}
 
The LXe is excited by the 570 and 1064 keV $\gamma$'s (and their
associated internal conversion electrons and X-rays) emitted by a $^{207}$Bi
source.  The source is prepared as a monoatomic layer electro-plated
on a $\simeq 3\mu$m diameter carbon fiber that is then threaded through the
cathode grid.  The thin source minimizes the effects of self-shadowing
of the scintillation light and is essential to obtain acceptable
scintillation resolution.  The data presented here were collected at
drift fields $E_{KG}$ of 0.2kV/cm, 0.5kV/cm, 1kV/cm, 2kV/cm, and
4kV/cm.

\begin{figure}[htb!!]
\begin{center}
\mbox{\epsfig{file=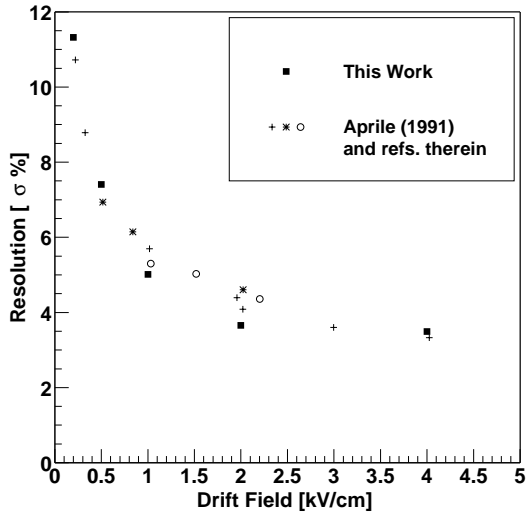,height=7.5cm}}
\vskip 0.1cm
\caption{Ionization resolution for 570~keV $\gamma$s in LXe when fit
  to a one-dimensional gaussian.  The black squares show the resolution 
  observed by this experiment for different electric fields. The other 
  symbols represent the resolutions observed with LXe ionization 
  chambers by other authors~\cite{resolutions}, in good agreement 
  with our data.
  }
\label{fig:resolution}
\end{center}
\end{figure}

A two-dimensional scatter plot of ionization versus scintillation
signals obtained at 4~kV/cm is shown in Figure~\ref{fig:4kv_correl}.
The two peaks in the ionization channel can be fit to a gaussian
function plus a first order polynomial function.   The energy
resolution obtained from the ionization data, once the electronics 
noise is subtracted in quadrature, is in good quantitative agreement 
with the results of other authors, as shown for the 570~keV peak in 
Figure~\ref{fig:resolution}.  

As shown in Figure~\ref{fig:4kv_correl}, the scintillation spectrum
has substantially poorer resolution and a larger continuum than the
ionization spectrum.     Differences between the two channels arise 
from the fact that the VUV photon acceptance depends somewhat upon 
the position of the energy deposited, as mentioned above, and
from the different fiducial volumes.  The resolution in the scintillation 
channel, as extracted by a fit with a gaussian plus a first degree 
polynomial, is $\sigma = (16.5 \pm 2.5)\%$ at 570~keV, where the error 
refers to the range of values observed at different fields.

The observed ionization and scintillation spectra are qualitatively
well reproduced in both peak positions and strengths by a detailed
GEANT~4~\cite{GEANT4} simulation which includes the processes of
internal capture, X-ray and Auger electron emission.  An empirical
gaussian smearing is used to match the
GEANT~4 output with the data.  However, in the absence of a microscopic 
model for the production of scintillation photons in LXe, the 
simulation could not be used to quantitatively fit the  
two dimensional spectrum. 

\begin{figure}[htb!!!!!!!!!!!!!!!!!!]
\begin{center}
\mbox{\epsfig{file=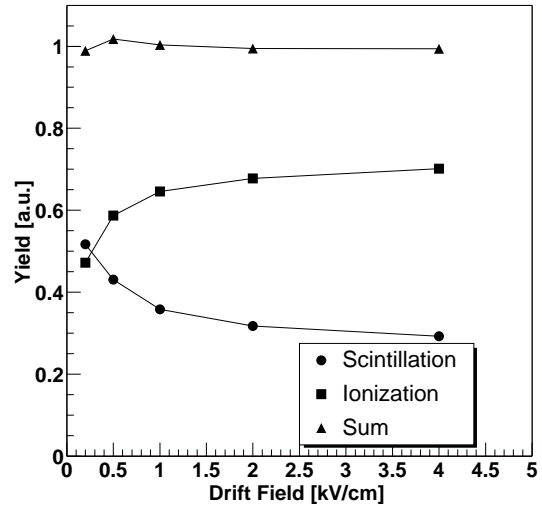,height=7.5cm}}
\vskip 0.1cm
\caption{Macroscopic anti-correlation between scintillation light and ionization
as a function of the electric field.   Similar results have been obtained before 
by other workers.}
\label{fig:MacroCorr1}
\end{center}
\end{figure}

Previous works~\cite{macro_corr} have reported evidence for what we call here 
``macroscopic'' anti-correlation between scintillation light and ionization 
as a function of the electric field applied to the LXe.  This effect, 
illustrated for our data in Figure~\ref{fig:MacroCorr1}, is thought to be due 
to the fact that scintillation occurs as ionization recombines. Different
values of the electric field modify such recombination, changing the proportion 
between light and ionization but leaving their sum constant.   
The constancy of the sum is 
well reproduced by data, as shown, in our case, by Figure~\ref{fig:MacroCorr1}.
We note that the absolute value of the sum does not directly correspond to
the total energy deposited, as part of the energy is lost in other channels.
We refer to this anti-correlation as ``macroscopic'' because is it an average
behavior of the LXe and it is obtained by changing an external, macroscopic,
parameter (the electric field).

In Figure~\ref{fig:4kv_correl}, clear ``microscopic'' anti-correlation is 
evident in each of the two ``islands'' corresponding to the 570~keV and 
1064~keV $\gamma$ lines.  This effect,
simply referred to as anti-correlation below, indicates that
energy sharing between scintillation and ionization fluctuates event by event
even though the sum of the two remains constant, in analogy with the macroscopic
correlations described above.
The microscopic nature of the anti-correlation may help in understanding the
large Fano factor mentioned above.   It also shows the way to dramatic improvements
in the energy resolution of  particle detectors, since the fluctuations clearly
affect scintillation and ionization in opposite ways.
Anti-correlations similar to the ones shown in Figure~\ref{fig:4kv_correl} 
are obtained at all fields and were not previously observed in LXe, although
they were shown, without a detailed discussion, for heavy ions in liquid 
Argon, by Ref.~\cite{Shibamura}.

The anti-correlation can be formally extracted from the data by
fitting each of the peaks to a 2-dimensional gaussian
distribution~\cite{stat1}

\begin{equation}
G(N_e,N_p)= H e^{ \large ( \frac{-1}{2 (1-\rho_{ep}^2)} \large [
\frac{\Delta N_e^2}{\sigma_e^{\prime 2}}+ \frac{\Delta
N_p^2}{\sigma_p^2} - \frac{ 2 \rho_{ep} \Delta N_e \Delta
N_p}{\sigma^{\prime}_e \sigma_p } \large ] \large )}\label{eqn:2Dfit}
\end{equation}

\noindent where $H = 1 / 2 \pi \sigma^{\prime}_e \sigma_p \sqrt{1 -
\rho_{ep}^2}$ is a normalization constant, $\rho_{ep} = V_{ep}/
\sigma^{\prime}_e \sigma_p $ is the correlation coefficient between
ionization and scintillation signals, $V_{ep} = \langle (N_e -
\bar{N_e})( N_p - \bar{N_p} ) \rangle $ is the covariance, $\Delta N_e
= N_e - \bar N_e$ and $\Delta N_p = N_p - \bar N_p$.  While the
instrumental noise on the scintillation channel is negligible and
hence ignored, the variance for the ionization channel, $\sigma^{\prime
2}_e = \sigma^2_e + \sigma^2_N$, includes the term, $\sigma_N = 381~e$,
that accounts for the readout noise mentioned above.    This is a relatively
small correction (e.g. 12\% at 4~kV/cm for the 570~keV peak).  Unlike the
one-dimensional fits used for ionization and scintillation data separately, 
the 2-dimensional fitting function does not include a background term.
A two-step procedure is used to first find an elliptical region of 
full width $2.35 \sigma$ around each of the two $\gamma$-ray peaks and then 
perform the final fit.
The limited range of the fit
suppresses the K-shell internal capture electrons and Compton
scattering events.  We found that the two-dimensional gaussian fits
tend to produce larger variances with respect to the one-dimensional
fits, providing a conservative estimate of the real resolutions. 

\begin{figure}[htb!!]
\begin{center}
\mbox{\epsfig{file=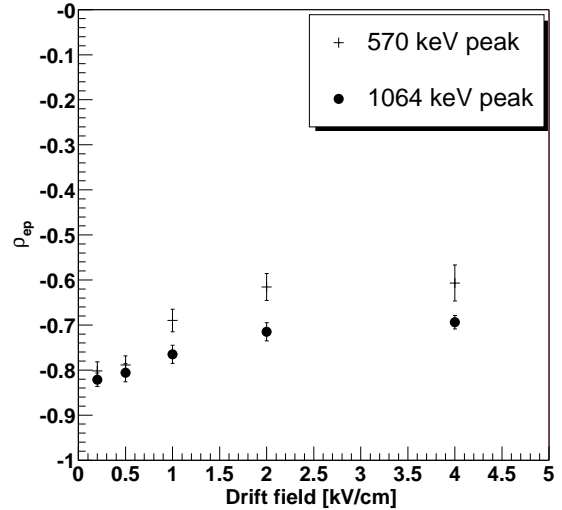,height=7.5cm}}
\vskip 0.1cm
\caption{Correlation coefficients between ionization and scintillation
signals obtained from the 2-dimensional fit discussed in the text.
The two sets of points refer to the coefficients found for the 570 and
the 1064~keV peaks.    The errors shown are from systematics in the analysis,
while statistical errors are negligible.}
\label{fig:CC_new}
\end{center}
\end{figure}

\begin{figure}[htb!!]
\begin{center}
\mbox{\epsfig{file=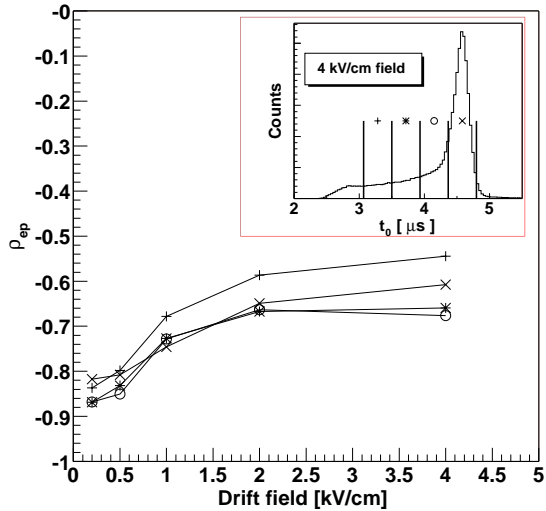,height=7.5cm}}
\vskip 0.1cm
\caption{Correlation coefficient at all fields for different subset of events at a 
4kV/cm field.
The complete data set for the 570~keV region, is divided in four regions, according 
to their drift-times, as illustrated by the effective drift-time distribution in 
the inset.   The peak to the right of this distribution is mostly due to electrons 
that stop near the source on the cathode.   For each field the cuts defining different 
regions are set in such a way as to select equivalent populations of events.
The magnitude of the correlation coefficients and their behavior as function
of the electric field do not depend significantly upon the energy deposition location.}
\label{fig:OffsetCutSummary}
\end{center}
\end{figure}

The correlation coefficients found by the fit are plotted in
Figure~\ref{fig:CC_new} as a function of drift field.  A clear
anti-correlation exists at all fields for both the 570~keV and the
1064~keV lines.    The errors shown in the Figure are systematic and
are estimated as the full range of the variation of $\rho_{ep}$ when
the different analysis cuts and fit intervals are changed.   
Statistical errors are negligible.
The stability of the correlation with respect to different experimental 
circumstances (such as position in the chamber, size of the chamber and
type of radiation) can be tested by repeating the fits for different 
subset of the data, each presenting a different range of effective 
drift-times, as derived from Eq.~\ref{eqn:fitFunction}.   
The effective drift-time 
distribution, shown for 4~kV/cm in the inset of 
Figure~\ref{fig:OffsetCutSummary}, allows for the selection of events
originating at different spatial positions along the drift field.
The peak at large drift-times primarily corresponds to electrons that
lose their energy and stop in the vicinity of the cathode plane where the 
source is.    As shown in Figure~\ref{fig:OffsetCutSummary} the correlation 
coefficient does not depend significantly on different effective drift times.

\begin{figure}[tb!!]
\begin{center}
\mbox{\epsfig{file=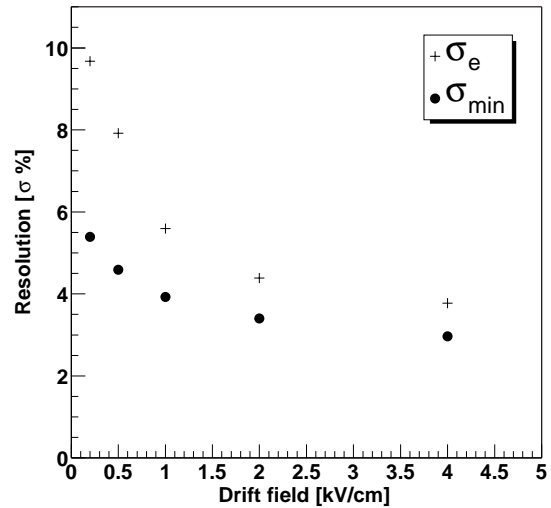,height=7.5cm}}
\vskip 0.1cm
\caption{Resolutions for the ionization channel and for the small
  correlated combinations using a fit to a two-dimensional correlated
  gaussian. The data shown is for the 570 keV gamma ray peak.}
\label{fig:Sigmas}
\end{center}
\end{figure}

By diagonalizing the
covariance matrix via a unitary rotation through angle $\theta = 2
\rho_{ep} \sigma_e \sigma_p/ (\sigma_e^2 - \sigma_p^2)$, the correlated
two-dimensional gaussian can be reduced to two uncorrelated gaussians
with variances $\sigma_{min}^2$ and $\sigma_{max}^2$.\cite{stat1} This
rotation is mathematically equivalent to measuring the variances along
the major and minor axes of the two-dimensional ellipses.  
Figure~\ref{fig:Sigmas} shows the ionization resolutions from the
2-dimensional fit and the minimum resolutions in the frame rotated by
$\theta$ for the 570 keV gamma ray peak at all drift fields.

The resolution improvement observed when correlating the ionization
and scintillation signals is significantly better than the uncorrelated
addition of the two spectra, where 
$\sigma_{p+e} = (\sigma_p^{-2} + \sigma_e^{-2})^{-1/2}$.    At 4~kV/cm
$\sigma_{p+e}=3.7\%$.   When using the correlated nature of the two channels 
the resolution $\sigma_{min}=3.0\%$, a factor of 1.24 improvement.
The improvement is more dramatic at the 0.2 kV/cm field where the
ratio is $9.7 / 5.4 = 1.8$.    While we found that the absolute 
values of the resolution change somewhat for different fitting methods 
and at different stages of the iterative procedure, the improvement factor 
is very stable.
We note here that the main results of this work are to show that a 
microscopic correlation exists and it can be consistently used to 
improve the resolution.   The ultimate resolution
will have to be measured at a later stage with a detector of larger 
volume, better VUV light collection and, possibly, a different source and
it is expected to be even better than the resolution $\sigma_{min}$ reported in 
Figure~\ref{fig:Sigmas}.

While the nature of the large Fano factor in liquefied noble gases 
is still not well understood, the different models considered generally involve
fluctuations in the density of the energy deposited~\cite{theory}.    Such fluctuations
can arise from density variations in the medium or from the small number 
of delta electrons with energies between 1 and 30 keV that are produced by 
Rutherford scattering of atomic electrons in the liquid.  
When these electrons come to a stop they produce a dense cluster of 
electron-ion pairs whose recombination rate is relatively high.  
In both cases more recombination occurs where the ionization density is higher.
Hence the statistics relevant for the fluctuations in the liquid is that
of the (small) number of high ionization density regions (either the number of 
delta-electrons or the number of high density regions).   In a sense, this would
mean that the Fano factors computed in Eq.~\ref{eqn:Fano} are too small because the
``wrong'' $N$ is used.

As first suggested in Ref.~\cite{seguinot} the
recombination hypothesis above implies that the fluctuations of ionization
and scintillation should be anti-correlated, since the high ionization-density
regions which produce less free charge should emit a correspondingly larger
amount of scintillation light.  A linear combination of ionization and
scintillation should therefore have smaller fluctuations than either
alone. Note that since the Fano factors $F_e$ and $F_p$ are typically
far in excess of their Poissonian values, the improvement should in
principle be substantial and is not limited by the factor of
$\sqrt{2}$ one would expect from the combination of two uncorrelated
signals of equal resolution. The data presented here provides direct
experimental evidence of this fluctuation and the associated recombination
by observing a clear microscopic anti-correlation between ionization 
and scintillation.
The correlation, $\rho_{ep}$, and the magnitude of the variances,
$\sigma_{max}^2$ and $\sigma_{min}^2$, provide input to improve the
model in terms of observables that are more directly related to the
microscopic physics.    It is interesting to note that from 
Figure~\ref{fig:CC_new} it appears that the magnitude of the correlation 
$\rho_{ep}$ decreases for increasing electric field (at least at low fields).
This behavior is possibly connected to the dependence of the intensity of 
the recombination luminescence with the electric field, as expected in all
the models discussed.

In conclusion, we have observed a clear microscopic anti-correlation
between ionization and scintillation in liquid xenon.
We have also measured the correlation coefficient between
these two channels to be $ -0.8 < \rho_{ep} < -0.6$.  
This observation has two important consequences. First, it provides 
direct experimental
evidence that recombination is a substantial source of the anomalously
large Fano factors in the high density liquid state.  Second, it
provides a method for improving the resolution of liquefied noble gas
calorimeters.  

One of us (G.G.) is indebted to P.~Picchi, F.~Pietropaolo and T.~Ypsilantis
for early discussions on the subject, M.~Moe for a careful reading of the 
manuscript and to R.G.H.~Robertson for having
stimulated the interest on this topic.
This work was supported by US-DoE grant DE-FG03-90ER40569-A019 and by
a Terman Fellowship of the Stanford University.


\end{document}